%% file: placement_paper.tex
\documentclass[journal, onecolumn]{IEEEtran}

\usepackage{tikz}
\usetikzlibrary{arrows,shapes}
\tikzset{
treenode/.style = {align=center, inner sep=0pt, text centered,
font=\sffamily},
}
\usepackage{epsfig}
\usepackage{pifont}
\usepackage{color}
\usepackage{graphicx}
\usepackage{bm}
\DeclareGraphicsExtensions{.png}
\usepackage{stfloats}
\usepackage{amsfonts}
\usepackage{amsbsy}
\usepackage{amsmath, amsthm, amssymb}
\usepackage{latexsym}
\usepackage{url}
\usepackage{cite}
\usepackage{color}
\usepackage{framed}
\usepackage{algorithm}
\usepackage{algpseudocode}
\usepackage{caption}
\usepackage{subcaption}
\usepackage{soul}
\usepackage{verbatim}

\DeclareMathOperator*{\argmax}{arg\,max}
\DeclareCaptionFormat{myformat}{#3}
\newtheorem{theorem}{Theorem}

\newtheorem{definition}{Definition}

\newtheorem{proposition}{Proposition}
\newtheorem{remark}{Remark}
\newtheorem{assumption}{Assumption}

\newcommand{\Fcal}{{\cal F}}
\newcommand{\Gcal}{{\cal G}}
\newcommand{\Hcal}{{\cal H}}

\newcommand{\Mcal}{{\cal M}}

\newcommand{\Pcal}{{\cal P}}

\newcommand{\Scal}{{\cal S}}

\newcommand{\lhat}{\hat{l}}

\newcommand\blfootnote[1]{%
  \begingroup
  \renewcommand\thefootnote{}\footnote{#1}%
  \addtocounter{footnote}{-1}%
  \endgroup
}

\begin{document}

\title{Sensor Placement for Outage Identifiability in Power Distribution Networks}
\author{Ananth Narayan Samudrala,~\IEEEmembership{Student Member,~IEEE},
M. Hadi Amini,~\IEEEmembership{Student Member,~IEEE},\\
        Soummya Kar,~\IEEEmembership{Member,~IEEE},
				        Rick S. Blum,~\IEEEmembership{Fellow,~IEEE}.
}

\maketitle
\begin{abstract}
Accurate topology information is critical for effective operation of power distribution networks. Line outages change the operational topology of a distribution network. Hence, outage detection is an important task. Power distribution networks are operated as radial trees and are recently adopting the integration of advanced sensors to monitor the network in real time. In this paper, a dynamic-programming-based minimum cost sensor placement solution is proposed for outage identifiability. We propose a novel formulation of the sensor placement as a cost optimization problem involving binary placement decisions, and then provide an algorithm based on dynamic programming to solve it in polynomial time. The advantage of the proposed placement strategy is that it incorporates various types of sensors, is independent of time varying load statistics, has a polynomial execution time and is cost effective. Numerical results illustrating the proposed sensor placement solution are presented for multiple feeder models including standard IEEE test feeders. 

\end{abstract}


\input{section1}

\input{section2}

\input{section3}

\input{section4}

\input{section5}

\input{section6}

\input{section7}

\bibliographystyle{IEEEtran}
\bibliography{References1}
\vfill

\end{document}

%% file: section1.tex
\section{Introduction}
\blfootnote{This work was supported by the Department of Energy under Award DE-OE0000779.

Ananth Narayan Samudrala and Rick S. Blum are with the Department of Electrical and Computer Engineering, Lehigh University, Bethlehem, PA 18015 USA (e-mail: \{ans416, rblum\}@lehigh.edu).

M. Hadi Amini and Soummya Kar are with the Department of Electrical and Computer Engineering, Carnegie Mellon University, Pittsburgh, PA 15213 USA (email: \{mamini1, soummyak\}@andrew.cmu.edu).}

Distribution network operators require accurate estimates of the network topology. Real time awareness of the operational topology of a distribution network is necessary for executing several important tasks \cite{monitoring_DER,sectionalization,demand_response,DSSE_baran}. However, line outages in a distribution network change the operational topology. Line outages are open lines in a distribution network that occur as a result of protective devices automatically isolating some part of the network. The cause for this isolation could be faults or physical topology attacks. A physical topology attack alters the physical dynamics of the network by physically removing bus interconnections \cite{topologyattackKar1}. Regardless of the reason for an outage, the isolation caused by it results in part of the network being disconnected from the main grid to form an island. Due to the loss of power supply from the main grid, the island might not be energized. Alternatively, with the advent of distributed generators (DGs) it is possible that the island remains energized by receiving power from a DG unit \cite{kati}. Regardless of the state of the island, real time detection of line outages (termed as outage detection) and the current operational topology of the network (termed  as topology identification) is an important task to return stability to the distribution network.   
\subsection{Related Work}
The radial structure of distribution networks has traditionally led them to have fewer monitoring devices. However, increasing load and consumption demands has increased the need for more reliable and secure operation of the grid. Therefore, distribution grid operators have recently started adopting smart grid technology \cite{sentient,linewatch,tollgrade} for grid monitoring. For example, Florida Power and Light Co. has signed a contract with Sentient Energy, Inc. ordering 20,000 line sensors\cite{florida}. Also, high precision phasor measurement units (PMUs) called micro-PMUs ($\mu$PMUs) are being designed specifically for distribution networks\cite{von}. A PMU is a device that estimates the magnitude and phase angle of an electrical phasor quantity like voltage using a common time source for synchronization \cite{pmu}. Assuming the widespread adoption of such sensors in distribution networks, new topology identification and outage detection methods are being proposed. Topology identification methods using optimal matching loop power\cite{gao}, time series measurements from PMUs \cite{time_series_dist}, data from smart-meters \cite{watson}, mixed integer quadratic programming\cite{tian} and voltage correlation data \cite{backhaus} were proposed previously. Utilizing power flow measurements from sensors along with load statistics, the authors of \cite{zhao,sevlian} propose maximum a-posteriori (MAP) and maximum likelihood (ML) outage detection algorithms respectively.

An important first step in employing sensors is their strategic placement in a network. Sensor placement techniques for state estimation \cite{muscas} and anomaly detection \cite{anna} were proposed previously. Specifically for outage detection, \cite{zhao,sevlian} propose sensor placement algorithms that support their corresponding outage detection algorithms and are optimal for them. However, the placement algorithms of \cite{zhao,sevlian}  have a few drawbacks. Firstly, they are dependent on load  statistics, i.e., expected load demands. Hence, the sensor placement solutions are only optimal for the expected load demands that were considered during the placement. Since expected load demands are time variant, this dependency might affect the optimality of the placement or worse could make outage detection of some line outages with the current placement impossible. Secondly, the placement algorithms consider all nodes in the network to have non-zero loads. However, most distribution networks have zero-injection nodes, due to which the proposed detection algorithms are unable to detect some outage scenarios. Finally, the placement algorithms are computationally intensive since they involve multiple computations of probability of missed detection error for various outage hypotheses. 

\subsection{Our Contribution}

{In this paper, we present a dynamic-programming-based minimum cost sensor placement solution for outage detection in distribution networks. It was shown in \cite{sevlian} that real power flow measurements are superior to voltage measurements when using only a small number of measurements for detecting outages in the distribution network. This is due to the small voltage differences between nodes which makes distinguishing changes between them difficult. Further, outages result in larger deviations in real power flows as compared with voltages. In order to deal with these issues, in our previous work \cite{ananth}, considering sensors that measure both real power flows and voltages, we have proposed a novel formulation of the sensor placement problem as a discrete cost optimization problem \cite{ananth}. In this paper, we expand and build on  our initial formulation in  \cite{ananth}. To this end, first, we provide systematic definitions of concepts called topology detectability and outage identifiability. These are important criteria that determine the capability of a sensor placement to provide measurements that are sufficient to distinguish and detect all outage scenarios. Next, we present the cost optimized sensor placement problem of \cite{ananth}. We further describe how the constraints of our optimization problem are necessary and sufficient for outage identifiability. Exploiting the radial structure of a distribution network, we then propose a dynamic programming algorithm that provides an optimum solution for our optimization problem in polynomial time. The advantage of our placement algorithm in comparison to those of \cite{zhao,sevlian} is that our solution considers various types of sensors, is independent of load statistics and line parameters, enables detection of line outages involving zero-injection nodes, and is computationally efficient}. 

The remainder of this paper is organized as follows. In section \ref{section2}, we present a description of the system model that would be employed in this paper. In section \ref{section3} we define topology detectability and outage identifiability. We describe our optimal sensor placement problem in section \ref{section4} and present the dynamic programming algorithm to solve it in section \ref{section5}. Section \ref{section6} presents numerical results and finally, conclusions are drawn in section \ref{section7}.

%% file: section2.tex
\section{System Model}\label{section2}

\subsection{Topology of a Distribution Network}

{We model the nominal (outage free) distribution network as a radial tree $\Gcal = \left\{V,E\right\}$ with $N$ nodes, where $V$ is the set of nodes and $E$ is the set of edges.} We have $V = \left\{1,2,\cdots,N\right\}$. Without loss of generality (WLOG) we consider node $1$ as the point of common coupling (PCC), i.e. the substation or the point where the distribution network under analysis is connected to the main grid and hence node $1$ is the root node of the tree $\Gcal$. {Regarding the root node we make the following assumptions.

\begin{assumption} \label{ass1}
We assume that the root node is the only power source in the distribution network.
\end{assumption}

\begin{assumption} \label{ass2}
The edge that connects the root node (PCC) to the main grid carries the power supply required for the entire distribution network and is therefore most likely to experience an outage. Hence, we assume that the operator directly monitors this line and therefore we do not consider sensor placement for this edge.
\end{assumption}
}

An edge between nodes $i$ and $j$ with $i$ as the parent node, i.e., power flows in the direction $i$ to $j$, is represented as $\left(i,j\right)$. Hence, $E$ is the set of all edges $(i,j)$ of the network. For every node $i$, $d_i$ is the degree of the node, $C_i$ is the set of all its children and $p_i$ is its parent node. For every node $i$, we call the edge that connects it to its parent node as the parent edge of node $i$ and similarly we call the edges that connect it to its children as the child edges of node $i$. Also, if an edge $(i,j)$ is on the path that connects another edge $(k,l)$ to the root node then edge $(i,j)$ is said to be upstream of edge $(k,l)$ and edge $(k,l)$ is said to be downstream of edge $(i,j)$. Fig. \ref{example} illustrates a distribution network tree with $N=9$ nodes. Node $1$ is the root node. We have $V = \left\{1,2,\cdots,9\right\}$. For illustration, the edge between nodes $1$ and $2$ is represented by $(1,2)$ in the figure. Considering node $3$ as an example, we have its degree $d_3 = 4$, the set of its children $C_3 = \left\{5,6,7\right\}$ and its parent node $p_3 = 1$. For node $3$, the edge $(1,3)$ is its parent edge, and the edges $(3,5),(3,6),(3,7)$ are its child edges. Also, we can say that edge $(1,3)$ is upstream of edge $(6,9)$ and edge $(6,9)$ is downstream of edge $(1,3)$. 

\subsection{Load Model} \label{load}

{Each non-root node $i$ in the network has a power consumption load $l_i$. The forecast of each load is $\lhat_i$ with error $e_i = l_i-\lhat_i$. Assuming that the errors are mutually independent normal random variables $e_i \sim N(0,\sigma_i^2)$, the true load can be modeled as a random variable distributed as $l_i \sim N(\lhat_i,\sigma_i^2)$ \cite{zhao}}. In the vector case we can write $\bm{l} \sim N(\bm{\lhat},\bm{\Sigma})$ where $\bm{l}$ and $\bm{\lhat}$  are the vector of true loads and vector of load forecasts at each node, and $\bm{\Sigma}$ is a diagonal covariance matrix. In any real world distribution network, there are at least a few nodes that do not have any load consumption, called zero-injection nodes. Representing the set of all zero-injection nodes in a network as $Z$, we have that zero-injection node $i \in Z$ has $l_i = \lhat_i = \sigma_i = 0$. {In Fig. \ref{example} we have no zero-injection nodes, i.e., $Z = \Phi$ where $\Phi$ represents an empty set.} Regarding the set $Z$ we make the following assumption.

\begin{assumption} \label{ass3}
We assume that the set $Z$ remains constant for all time, i.e., a zero-injection node $i \in Z$ will always have $\lhat_i = \sigma_i = 0$ and a nonzero-injection node $j \in V \setminus Z$ will always have $\lhat_j > 0$ and $\sigma_j \geq 0$. \end{assumption}

{
Regarding the loads at nonzero-injection nodes please note following remark.

\begin{remark}\label{rem1}
We do not utilize the distributions of load statistics of nonzero-injection nodes and values of line parameters of distribution network lines during sensor placement. Hence, they will not affect the optimality of our sensor placement. However, we consider that knowledge of current load statistics and line parameters is available for outage detection.
\end{remark}
}

\subsection{Sensor Types}

We broadly classify the various sensors available for distribution system monitoring into two categories: line sensors and node sensors, and consider placing these two types of sensors in our placement algorithm.

\emph{Line Sensor:}A line sensor is installed on an edge of a distribution network tree. Examples of line sensors are Sentient MM3 \cite{sentient} and Tollgrade Lighthouse MV\cite{tollgrade}. Line sensors vary in their measurement capabilities from manufacturer to manufacturer. In this paper we consider that the point of installation of a line sensor on an edge $(i,j)$ is towards the end of the edge connected to node $j$, and that the line sensor measures the real power flow on the edge and the voltage magnitude at node $j$. 

\emph{Node Sensor:} A node sensor is installed at a node of a distribution network tree. Examples of node sensors are Linewatch L \cite{linewatch} and micro-PMUs \cite{von}. Node sensors vary in their measurement capabilities from manufacturer to manufacturer. In this paper, we consider that a node sensor installed at a node measures the real power flow on every edge of the node, and the voltage magnitude at the node. 

\begin{figure}[h]
	\centering
	\begin{tikzpicture}[
	  -,
		level/.style={sibling distance = 3cm/#1,level distance =1.3cm},
    arn_r/.style = {treenode, circle, black, font=\sffamily\bfseries, draw=red, very thick,text width=2.0em},
    arn_w/.style = {treenode, circle, black, font=\sffamily\bfseries, draw=black, text width=2.0em, very thick},
		edg_b/.style = {edge from parent/.style = {black,thick,draw}},
		edg_r/.style = {edge from parent/.style = {green,very thick,draw}}] 
	\node [arn_r] {1}
	child[edg_b]{ node [arn_w] {2}
	  child[edg_b]{ node [arn_w] {4}}}
  child[edg_b]{ node [arn_w] {3}
	  child[edg_b]{ node [arn_w] {5}  
		  child[edg_b]{ node [arn_w] {8}}}
		child[edg_r]{ node [arn_w] {6}
		  child[edg_b]{ node [arn_w] {9}}}
		child[edg_r]{ node [arn_w] {7}}};
	\end{tikzpicture}
	\caption{A distribution network represented as a tree.}
	\label{example}
\end{figure}
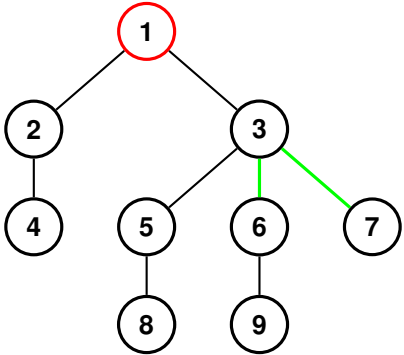

In Fig. \ref{example}, we have line sensors (in green) on edges $(3,6)$ and $(3,7)$, and a node sensor (a red circle) at node $1$. We can represent the sensor placement configuration for the network by $\Pcal = (V_{\Pcal},E_{\Pcal})$ where the set of nodes endowed with a node sensor is $V_{\Pcal} \subseteq V$ and the set of edges endowed with a line sensor is $E_{\Pcal} \subseteq E$. For Fig. \ref{example}, we have $V_{\Pcal} = \left\{1\right\}$, $E_{\Pcal} = \left\{(3,6),(3,7)\right\}$.

{
\subsection{Power Flow Model and Sensor Measurements}

{We consider the linearized DistFlow equations for the power flow model \cite{dobbe}. Since we consider that the sensors measure only real power flows, from now on we shall simply refer to real power flows as power flows. Under assumption \ref{ass1}, according to the linearized DistFlow equations we can write the true power flow $\hat{s}_{(i,j)}$ on an edge $(i,j)$ as the sum of all loads downstream of that edge, i.e., $\hat{s}_{(i,j)} = \sum_{k \in T_j}l_k$ where $T_j$ is set of all nodes in the sub-tree rooted at node $j$. 

Coming to the sensor measurements}, let $\Scal_{\Pcal}$ be the set of all edges whose power flow is measured either by a node or a line sensor or both under the placement $\Pcal$. Let $\Mcal_{\Pcal}$ be the set of all nodes which have their voltage magnitude measured either by a node or a line sensor or both. The power flow measurement on an edge $(i,j) \in \Scal_{\Pcal}$ is represented as $s_{(i,j)}$. We have

\begin{equation}\label{e1}
    s_{(i,j)} = \hat{s}_{(i,j)} + n_{(i,j)} \ \forall (i,j) \in \Scal_{\Pcal}
\end{equation}
where $n_{(i,j)}$ is the sensor noise. For example, in Fig. \ref{example} we have $\hat{s}_{(3,6)} = \sum_{k \in T_6}l_k = l_6+l_9$ since $T_6 = \{6,9\}$. Since $l_k = \lhat_k + e_k$, we can re-write (\ref{e1}) as

\begin{equation}\label{e2}
    s_{(i,j)} = \sum_{k \in T_j} \lhat_k + \sum_{k \in T_j} e_k + n_{(i,j)} \ \forall (i,j) \in \Scal_{\Pcal}.
\end{equation}

Similarly, the voltage magnitude measurement at a node $j \in \Mcal_{\Pcal}$ is represented as $v_j$. We have
\begin{equation}\label{e3}
    v_j = \hat{v}_j + n_j \ \forall j \in \Mcal_{\Pcal}
\end{equation}
where $\hat{v}_{j}$ is the true voltage magnitude at node $j$ and $n_j$ is the sensor noise. 
}
\subsection{Outage Hypotheses}

We model outages as disconnected edges. An edge outage disconnects a part of the network by breaking the network into two: an energized tree connected to the root node and an island with no power. For example, outage of edge $(3,6)$ in Fig. \ref{example} disconnects nodes $6$ and $9$ from the main network forming an island with no power. We define an outage hypothesis $H$ as the set of all edges in outage, i.e., $H = \left\{(i,j) \in E| \ \text{Edge} \ (i,j) \ \text{is in outage}\right\}$. Let the set of all outage hypotheses of $\Gcal$ be $\Hcal$ with $|\Hcal| = 2^{N-1}$ where $|\Hcal|$ represents the number of elements in set $\Hcal$. Every line outage results in the formation of one new island, so an outage hypothesis $H$ results in a forest of trees $\Fcal^H = \left\{\Gcal^H_0,\Gcal^H_1,\cdots,\Gcal^H_{|H|}\right\}$. WLOG we represent the energized tree by $\Gcal^H_0$. As an example consider Fig. \ref{example} with the outage hypothesis $H_1 = \left\{(3,6)\right\}$. Under $H_1$, we have $\Fcal^{H_1} = \left\{\Gcal^{H_1}_0,\Gcal^{H_1}_1\right\}$. The remaining energized tree is $\Gcal^{H_1}_0 = (V^{H_1}_0,E^{H_1}_0)$ with $V^{H_1}_0 = \left\{1,2,3,4,5,7,8\right\}$ and $E^{H_1}_0 = \left\{(1,2),(1,3),(2,4),(3,5),(5,8),(3,7)\right\}$. The disconnected island is $\Gcal^{H_1}_1 = (V^{H_1}_1,E^{H_1}_1)$ with $V^{H_1}_1 = \left\{6,9\right\}$ and $E^{H_1}_1 = \left\{(6,9)\right\}$.

%% file: section3.tex
{\section{Topology Detectability and Outage Identifiability}\label{section3}

In this section we define the terms: topology detectability and outage identifiability, that are important in the context of sensor placement for outage detection. {In defining the two terms, in addition to assumptions \ref{ass1}-\ref{ass3} we make the following assumption.}

\begin{assumption} \label{ass4}
We assume that the sensor measurements obtained from a sensor placement $\Pcal$ are noise free, i.e., $n_{(i,j)} = 0 \ \forall (i,j) \in \Scal_{\Pcal}$ and $n_j = 0 \ \forall j \in \Mcal_{\Pcal}$, and that load forecasting is perfect, i.e., $e_i = 0 \ \forall i \in V \setminus \{1\}$.  
\end{assumption}
{
The following remark discusses the effect of assumption \ref{ass4} on performance of detection algorithms.

\begin{remark}\label{rem2}
Since detectability and identifiability will be defined under noise free sensing, the detection performance of any algorithm that is used for detecting the topology or outage hypothesis, will be binary, i.e., either the algorithm will detect the correct topology or outage hypothesis with absolute certainty, or it will be unable to detect.  
\end{remark}

From remark \ref{rem2} we can see that topology detectability and outage identifiability would depend only on the structure of a distribution network. The following definition formalizes the concept of topology detectability.

\begin{definition}[Topology Detectability]\label{def1}
A distribution network $\Gcal$ is said to be topology detectable using a sensor placement $\Pcal$, {if under assumptions \ref{ass1}-\ref{ass4}} there exists a topology detection algorithm that can detect the topology of the energized tree $\Gcal^H_0$ under every outage hypothesis $H \in \Hcal$ by utilizing the measurements obtained from the placement $\Pcal$ and available load statistics. The sensor placement $\Pcal$ is said to ensure topology detectability for the network $\Gcal$.
\end{definition}
}
The following definition describes the concept of a subset placement.

\begin{definition}[Subset Placement]\label{def2}
A sensor placement $\Pcal_1$ is a subset of a sensor placement $\Pcal_2$ represented as $\Pcal_1 \subseteq \Pcal_2$ if and only if $\Mcal_{\Pcal_1} \subseteq \Mcal_{\Pcal_2}$ and $\Scal_{\Pcal_1} \subseteq \Scal_{\Pcal_2}$.
\end{definition}

Topology detectability has the property of monotonicity which is stated in the following theorem.

\begin{theorem}\label{th1}
Let $\Pcal_1$ and $\Pcal_2$ be two different sensor placements for a network $\Gcal$ such that $\Pcal_1 \subseteq \Pcal_2$. Then, topology detectability of $\Gcal$ under $\Pcal_1$ implies topology detectability of $\Gcal$ under $\Pcal_2$. Conversely, lack of topology detectability of $\Gcal$ under $\Pcal_2$ implies lack of topology detectability of $\Gcal$ under $\Pcal_1$. 
\end{theorem}

Proof of theorem \ref{th1} is simple. Any algorithm that enables topology detectability under the placement $\Pcal_1$ can do the same under $\Pcal_2$ since $\Pcal_1 \subseteq \Pcal_2$, i.e., $\Pcal_2$ provides all measurements obtained from $\Pcal_1$ and some more. From this property, we can see that a sensor placement in which the power flows on all edges of a network and voltage magnitudes at every node of a network are measured by sensors, would ensure topology detectability for any network. Such a placement is termed as a maximal sensor placement $\Pcal_{max}$. Two examples of $\Pcal_{max}$ are a placement with node sensors at every node and a placement with a node sensor at the root node and line sensors on every edge. It is important to note that detecting the topology of the energized tree $\Gcal^H_0$ is different from identifying the exact outages that resulted in the topology change, i.e., identifying the set $H \in \Hcal$. For example, consider two outage hypotheses: $H_1 = \left\{(3,6)\right\}$ and $H_2 = \left\{(3,6),(6,9)\right\}$. Under both $H_1$ and $H_2$ the energized tree is the same, i.e., $\Gcal^{H_2}_0 = \Gcal^{H_1}_0$. Topology detectability refers only to detecting the topology of the energized tree, i.e., $\Gcal^{H_1}_0$ or $\Gcal^{H_2}_0$, and does not refer to identifying the outage hypothesis $H_1$ or $H_2$ that caused it.  

Unfortunately, identification of the outage hypothesis $H \in \Hcal$ that resulted in an energized tree $\Gcal^{H}_0$ is not always possible, even with a maximal sensor placement. As an example consider the distribution network of Fig. \ref{example} but with a maximal sensor placement of node sensors at every node. With this maximal placement, consider again the two outage hypotheses $H_1$ and $H_2$. Since $(3,6)$ is in outage in both the outage hypotheses, all the measurements made by the node sensors would be exactly the same in both the outage situations making them indistinguishable. This is because outage edge $(6,9)$ is downstream of outage edge $(3,6)$. As the two outage hypotheses cannot be distinguished with a maximal placement, they cannot be distinguished by any placement. Since, all edges that are downstream of outages will be disconnected from the energized tree irrespective of whether they themselves are in outage or not, and all measurements will be exactly the same either way, outage detection is restricted to a set $\Hcal_U \subseteq \Hcal$ \cite{sevlian}. $\Hcal_U$ is the set of all uniquely identifiable outages, i.e., $\Hcal_U = \left\{H \in \Hcal | \ \text{No edge in outage is downstream of another}\right\}$. Considering outage detection over the set $\Hcal_U$ we define outage identifiability as follows.
{
\begin{definition}[Outage identifiability]\label{def1}
A distribution network $\Gcal$ is said to be outage identifiable using a sensor placement $\Pcal$, if under assumptions \ref{ass1}-\ref{ass4} there exists an outage detection algorithm that can identify every outage hypothesis $H \in \Hcal_U$ by utilizing the measurements obtained from the placement $\Pcal$ and available load statistics. The sensor placement $\Pcal$ is said to ensure outage identifiability for the network $\Gcal$.
\end{definition}
}
Similar to topology detectability, outage identifiability also has the property of monotonicity of Theorem 1. A maximum sensor placement $\Pcal_{max}$ will always ensure outage identifiability. Ensuring outage identifiability ensures topology detectability since detecting outages will lead to detection of the energized tree topology. Since, a maximal placement would not be cost effective, in this paper we propose a cost optimal sensor placement solution that ensures outage identifiability. It is important to note here that topology detectability and outage identifiability are defined under the noise free conditions of assumption \ref{ass4}. However, in practice sensor measurements and load forecasts will be noisy. But any sensor placement that ensures outage identifiability will be sufficient to work with an appropriate outage detection algorithm to detect outages in a noisy scenario, albeit with a performance limit that is determined by the noise in the measurements and load forecasts. Our sensor placement solution can provide critical sensing for statistical outage detection algorithm such as the ones proposed in \cite{zhao,sevlian}. However, the outage detection algorithms of \cite{zhao,sevlian} do not use voltage measurements and hence are unable to detect all outage scenarios of edges of zero-injection nodes. To address this issue, in a subsequent work we would be presenting an outage detection algorithm that can detect all edge outages in a distributed manner using noisy measurements obtained from our sensor placement and noisy load forecasts. 

}

%% file: section4.tex
\section{Optimal Sensor Placement}\label{section4}

In this section we formulate our sensor placement solution as an optimization problem that ensures outage identifiability at minimal cost. Let $x_i$ be a binary variable that determines if node $i \in V$ is endowed with a node sensor ($x_i = 1$) or not ($x_i = 0$). Let $\bm{x}$ represent the vector of all $x_i$. Similarly, let the binary variable $y_{(i,j)}$ determine if an edge $(i,j) \in E$ is endowed with a line sensor ($y_{(i,j)}=1$) or not ($y_{(i,j)}=0$). We have $\bm{y}$ as the vector all $y_{(i,j)}$. Let $a_i$ be the cost of placing a node sensor at node $i \in V$ and $b_{(i,j)}$ be the cost of placing a line sensor on edge $(i,j) \in E$. We assume $a_i \geq 0 \ \forall \ i \in V$ and $b_{(i,j)} \geq 0 \ \forall \ (i,j) \in E$. Then, the sensor placement problem is formulated as the following cost minimization problem $OP$.

\begin{align}
OP: \underset{\bm{x},\bm{y}}{\text{minimize:}}& \sum\limits_{i \in V} a_i x_i + \sum\limits_{(i,j) \in E} b_{(i,j)} y_{(i,j)} \label{obj} \\
\text{subject to:}
& \, d_1 x_1 + \! \sum\limits_{j \in C_1} x_j + \! \sum\limits_{(1,j) \in E} y_{(1,j)} \geq d_1-1 \label{c1}\\
& d_k x_k + \! \sum\limits_{j \in C_k} x_j + \! \sum\limits_{(k,j) \in E} y_{(k,j)} \nonumber \\
						 &\geq d_k-2 \ \forall k \in V \setminus \left\{1\right\} \ \text{having} \ d_k \geq 3 \label{c2} \\
& x_k + y_{(p_k,k)} \geq 1 \ \forall k \in Z \label{c3}. 
\end{align}
{
The following proposition discusses the necessity of constraints of $OP$ in ensuring outage identifiability. 
\begin{proposition}\label{prop1}
 Under the linearized DistFlow power flow model, assumptions \ref{ass1}-\ref{ass4} and any possible load statistics, the constraints of $OP$ are necessary and sufficient for an outage detection algorithm to identify the correct outage hypothesis $H \in \Hcal_U$. 
\end{proposition}

Now we shall explain how the constraints of $OP$ are necessary and sufficient for outage identifiability.} Firstly, constraint (\ref{c1}) ensures that the outages of child edges of the root node are identifiable. Since, the root node supplies power to all loads in a distribution network through its child edges, monitoring these edges directly with sensors is necessary to know if they are in outage (zero power flow) or not (non-zero power flow). The root node has $d_1-1$ child edges since it has one parent edge that connects it to the main power grid. Hence, constraint (\ref{c1}) requires that a combination of a node  and line sensors must monitor all $d_1-1$ child edges of the root node. {In Fig. \ref{example} the child edges of the root node are monitored by a node sensor at the root node. To illustrate the necessity of constraint (\ref{c1}) assume that the constraint is violated by having a node sensor at node 4 instead of node 1. In that case,  under both an outage of edge $(1,2)$ or an outage of edge $(2,4)$, the node sensor at node 4 will make the exact same set of measurements, making it impossible for any algorithm to differentiate the two outage scenarios.}

Constraint (\ref{c2}) is for identifiability of outages of child edges of a non-root nodes with degree greater than or equal to $3$. A non-root node $k \in V$ with $d_k \geq 3$ has one parent edge and $d_k-1$ child edges. Since the parent edge is a child edge of another node, we can assume that the outages of the parent edge can be identified by a sensor upstream. Hence, we only need to ensure the identifiability of outages of the child edges. To achieve this we need to monitor at least $d_k-2$ child edges. Hence for a non-root node $k \in V$ with $d_k \geq 3$, constraint (\ref{c2}) requires that a combination of sensors must monitor at least $d_k-2$ child edges of node $k$. In Fig. \ref{example}, two child edges $(3,6)$ and $(3,7)$ of the non-root node $3$ are monitored by line sensors. Noise free power flow measurement of edge $(3,6)$ and perfect load forecast together enable identification of outages of edges $(3,6)$ and $(6,9)$. Since power flows are additive, power flow on edge $(3,5)$ can be obtained by subtracting the power flows of edges $(3,6)$ and $(3,7)$ from the power flow on edge $(1,3)$. Finally, outages of edge $(5,8)$ can be identified by using the computed power flow on edge $(3,5)$ and perfect load forecasts. {To illustrate the necessity of constraint (\ref{c2}) assume that we do not have a line sensor on edge $(3,6)$. In that case, it is possible that a certain combination of load statistics and line parameters might result in the remaining two sensors to make the exact same measurements under two different outage scenarios: outage of edge $(3,5)$ and outage of edge $(3,6)$. Hence, constraint (\ref{c2}) is necessary.} It is important to note here that outage identifiability of child edges of non-root nodes with degree $2$ also requires constraint (\ref{c2}). However, for a degree $2$ node the right hand side of (\ref{c2}) is zero and hence we need not place any sensors to satisfy (\ref{c2}). 

In addition to constraints (\ref{c1}) and (\ref{c2}) we need (\ref{c3}) for identifiability of edge outages of zero-injection nodes. Assume that node $3$ in Fig.\ref{example} is a zero-injection node. Consider two outage hypotheses $H_3 = \{(1,3)\}$ and $H_4 = \{(3,5),(3,6),(3,7)\}$. Under both $H_3$ and $H_4$, the node sensor at node 1 would measure zero flow on edge $(1,3)$ and the line sensors on edges $(3,6)$ and $(3,7)$ would measure zero power flow and zero voltage. Hence, sensor measurements from the sensor placement of Fig. \ref{example} cannot help distinguish $H_3$ and $H_4$. Constraint (\ref{c3}) resolves this issue by having a line sensor on edge $(1,3)$ or a node sensor at $3$. A line sensor on edge $(1,3)$ would measure zero voltage under $H_3$ but would give a non-zero voltage measurement under $H_4$. The same ideas are true if we had a node sensor at node 3. Thus the two scenarios can be distinguished. {In this way, the three constraints together are necessary to ensure that sufficient measurement data is available to distinguish and detect every outage hypothesis $H \in \Hcal_U$. Multiple hypothesis testing is an outage detection algorithm that works with our sensor placement. Since our sensor placement solution ensures outage identifiability, it can be employed for outage detection in practical scenarios where sensor measurements and load statistics are noisy.} The costs $a_i \forall i \in V$ and $b_{(i,j)} \forall (i,j) \in E$ in the objective function (\ref{obj}) are user defined and can vary depending on the sensor products and utility practices. {It is important to note that $OP$ does not consider load statistics anywhere. However, considering load statistics in sensor placement as in \cite{sevlian} reduces the number of sensors required, albeit with the disadvantage of dependency on load statistics.} In the following section, we propose a dynamic-programming-based algorithm that can provide an optimal solution to $OP$ in polynomial time. 

%% file: section5.tex
\section{Dynamic Programming Algorithm}\label{section5}

Dynamic programming is a technique that is used to solve many optimization problems in polynomial time for which a naive approach would take exponential time. For dynamic programming to be applicable, an optimization problem must have two key attributes: optimal substructure and overlapping sub-problems. Optimal substructure means that the solution to a given optimization problem can be obtained by a combination of optimal solutions to its smaller but similar sub-problems (Bellman's principle) \cite{cormen}. Overlapping sub-problems means that any recursive algorithm solving the problem should solve the same sub-problems over and over, rather than generating new sub-problems. To solve by dynamic programming we first need to define a sub-problem that has an optimal substructure and then remember the solutions to the sub-problems that we have already solved and re-use them (memoization). In our dynamic programming algorithm, we shall use the following notations. Firstly, we define a critical node set $V_{eval}$ as a set containing the root node, nodes with a minimum degree $3$ and zero-injection nodes, i.e., $V_{eval} = \left\{i \in V| \ i = 1 \ \text{or} \ d_i \geq 3 \ \text{or} \ i \in Z\right\}$. The depth of a node $i \in V$, represented by $f_i$, is the number of edges in the path that connects node $i$ to the root node. The depth of the root node is $0$, i.e. , $f_1 = 0$. The depth of the tree is the maximum of the depths of all nodes $i \in V$, i.e. , $f_{max} = \text{max}_{i \in V} f_i$.  Also, let $m_i = \text{max}(a_i,b_{(p_i,i)}) \forall i \in V-\left\{1\right\}$, i.e , $m_i$ is the maximum of the cost of node sensor at node $i$ and the cost of line sensor on parent edge of $i$.  

Our dynamic programming sub-problem is finding an optimal sensor placement for a sub-tree rooted at a node $i \in V_{eval}$. The sub-problem is solved by the function \textbf{Placement}. With this sub-problem definition, we start our algorithm at depth $f_{max}-1$ and traverse up the tree to depth $f_1 = 0$, by decreasing the depth by $1$ after each iteration. The number of iterations is equal to $f_{max}$. We start at depth $f_{max}-1$ since all the nodes at depth $f_{max}$ are nonzero-injection nodes with degree $2$ and therefore do not affect the constraints of $OP$. At each iteration $k$ the algorithm is at a search depth $f^{Alg}_k = f_{max}-k$. At each depth $f^{Alg}_k$, we make a list of all nodes at that depth that are in $V_{eval}$. Let this list be represented as $Q_{k} = \left\{i \in V_{eval}| f_i = f^{Alg}_k \right\}$. At each iteration $k$, for each node $q \in Q_k$ we execute the function \textbf{Placement}, i.e., we solve the sub-problem at $q$. If $q \in Q_k$ has $d_q \geq 3$ we have two competing solutions to the sub-problem at $q$: the optimal sensor placement could include a node sensor at $q$ or could include a combination of node and line sensors on it's children and child edges. We need to evaluate the cost for both the choices and choose the minimum. Similarly if $q \in Q_k$ is a zero-injection node we have two competing solutions to the sub-problem at $q$: the optimal sensor placement could include a node sensor at $i$ or a line sensor on parent edge of $i$. Again, we need to evaluate the cost for both the choices and choose the minimum. After solving the sub-problem at $q$, the function \textbf{Placement} updates the current placement $\Pcal$. This is equivalent to memoization. This process is repeated until the sub-problem at root node $1$ is solved to obtain an optimal solution to $OP$.  At the start of algorithm we have the placement as $\Pcal = (V_{\Pcal} = \Phi,E_{\Pcal} = \Phi)$ which is updated at every execution of the function \textbf{Placement}. The placement $\Pcal$ after the final iteration $k = f_{max}$ is the optimal placement. The function \textbf{Placement} employs the functions \textbf{NoSensorChild}, \textbf{Length} and \textbf{FindMax}. A brief description of these three functions is given below.

\begin{enumerate}
	\item \textbf{NoSensorChild}($q,\Pcal$): The inputs to this function are each node $q$ in each $Q_k$ and the current placement $\Pcal$. The function returns the set \textit{list} which is the set of child nodes of node $q$ that do not have a node sensor and whose parent edge does not have a line sensor. Hence, $\textit{list} = \left\{i \in C_q| i \notin V_{\Pcal}, (q,i) \notin E_{\Pcal}\right\}$.
	\item \textbf{Length}(\textit{list}): This function returns the number of elements in the input \textit{list}.
	\item \textbf{FindMax}(\textit{list}): This function has input \textit{list} and outputs the set $\textit{list2} = \textit{list}-\left\{\textit{maxnode}\right\}$ where $\textit{maxnode} = \argmax_{i \in \textit{list}} m_i$. $\textit{maxnode}$ is a single node. If multiple nodes in \textit{list} have the same $m_i$, then $\textit{maxnode}$ is the one with the least degree $d_i$. 
\end{enumerate}

\begin{algorithm}[!t] 
	\caption{{\bf Function:} \textbf{Placement}$(q \in Q_k,\ \Pcal)$}
	\begin{algorithmic}[1]
		\State{\bf{Input:}} Node $q \in Q_k$, current placement $\Pcal$.
		\State{\bf{Output:}} Updated $\Pcal$. 
		\State{\bf{Begin}}
		\State \textit{list} = \textbf{NoSensorChild}($q,\Pcal$)
		\If {$q == 1$ \& \textbf{Length}(\textit{list}) $> 0$ \& $a_q \leq \sum_{i \in \textit{list}}m_i$}
		         \State $V_{\Pcal} = V_{\Pcal} \cup \left\{1\right\}$
		\ElsIf {$q == 1$} 
		          \State $V_{\Pcal} = V_{\Pcal} \cup \left\{i \in \textit{list }| \ m_i == a_i\right\}$ 
		          \State $\textit{t} = \left\{(q,i) \in E| i \in \textit{list } \text{and } m_i == b_{(q,i)}\right\}$
		          \State $E_{\Pcal} = E_{\Pcal} \cup \textit{t}$ 
		\ElsIf {$q \in Z$ \& \textbf{Length}(\textit{list})
		$\leq 1$}
		     \If {$a_q \leq b_{(p_q,q)}$} 
			      \State $V_{\Pcal} = V_{\Pcal} \cup q$
		     \Else
			      \State $E_{\Pcal} = E_{\Pcal} \cup (p_q,q)$
			 \EndIf     
	    \Else
	          \State \textit{list2} = \textbf{FindMax}(\textit{list})
	          \If {$q \in Z$ \& $a_q \leq b_{(p_q,q)}+\sum_{i \in \textit{list2}}m_i$}
	               \State $V_{\Pcal} = V_{\Pcal} \cup \left\{q\right\}$
	          \ElsIf{$q \notin Z$ \& $a_q \leq \sum_{i \in \textit{list2}}m_i$}
                   \State $V_{\Pcal} = V_{\Pcal} \cup \left\{q\right\}$
	          \Else
	          	    \State $V_{\Pcal} = V_{\Pcal} \cup \left\{i \in \textit{list2 }| \ m_i == a_i\right\}$ 
	          	    \State $\textit{t} = \left\{(q,i) \in E| i \in \textit{list2 } \text{and } m_i == b_{(q,i)}\right\}$
		              \State $E_{\Pcal} = E_{\Pcal} \cup \textit{t}$ 
	          \EndIf
    	\EndIf
		\end{algorithmic}
	\label{placement}
\end{algorithm}

As an example to understand the dynamic programming algorithm consider the distribution network of Fig. \ref{example}. There are no zero-injection nodes in this network. Let $a_i = 2 \forall \ i \in V$, $b_{(3,6)} = b_{(3,7)} = 0.3$ and $b_{(i,j)} = 1 \forall (i,j) \in E-\left\{(3,6),(3,7)\right\}$. Initially, we have the placement $\Pcal = (V_{\Pcal} = \Phi,E_{\Pcal} = \Phi)$. The depth of the tree $f_{max} = 3$. Hence, our algorithm has $3$ iterations and starts with iteration $k = 1$ at depth $f^{Alg}_1 = 2$. There are no critical nodes at depth $2$. Hence we move up to depth $f^{Alg}_2 = 1$ for iteration $k = 2$. Here, we have $Q_2 = \left\{3\right\}$ since $d_3 = 4$. For node $3$ we execute the function \textbf{Placement}. As we have $a_3 = 2$ but $b_{(3,6)}+b_{(3,7)} = 0.6$, \textbf{Placement} returns $\Pcal$ with updated $E_{\Pcal} = \left\{(3,6),(3,7)\right\}$. Then for the final iteration we move up to $f^{Alg}_3 = 0$. Here, we have $Q_3 = \left\{1\right\}$. For node $1$ the function \textbf{Placement} is executed. Since $a_1 = 2$ and $b_{(1,2)}+b_{(1,3)} = 2$, we could either place a node sensor at $1$ or line sensors on it's child edges. Both placements are equivalent in terms of total cost. We chose to place a node sensor at $1$. Then the final placement as illustrated in Fig. \ref{example} is $\Pcal = (V_{\Pcal},E_{\Pcal})$ where $V_{\Pcal} = \left\{1\right\}$ and $E_{\Pcal} = \left\{(3,6),(3,7)\right\}$. 

The run-time $T$ of any dynamic programming algorithm is given by $T  = (\text{Number of sub-problems} \times \text{Number of competing solutions per sub-problem} \times \text{time per sub-problem})$ \cite{cormen}. In our dynamic programming placement algorithm we have $|V_{eval}|$ number of sub-problems. Each sub-problem has two competing solutions. The time per sub-problem is $O(1)$ since we employ memoization and therefore have access to optimal solutions of sub-problems that have already been solved \cite{cormen}. Hence, the run-time $T$ of our placement algorithm is $O(|V_{eval}| \times 2 \times 1)$, i.e. , $O(|V_{eval}|)$. Since $|V_{eval}| < N$, we have $O(|V_{eval}|) < O(N)$.

%% file: section6.tex
\section{Numerical Results} \label{section6} 

In this section, we present optimal sensor placement solutions for various distribution network models. We present three sets of numerical results. In the first set, we show that the optimum sensor placement is varies with costs. In the second set, we simulate sensor placement for a couple of standard IEEE test feeders, highlighting the change in optimal sensor placement with zero-injection nodes. Finally, we demonstrate the scalability of our dynamic programming algorithm. 

All results presented here were computed using MATLAB based code running on a 3.4GHz Intel(R) Core(TM) i7-2600 processor with 8 GB of RAM. In all results, an edge in green represents a line flow sensor and a node in red represents a node sensor. We used the MATLAB function \textit{timeit} to measure the algorithm execution times.

\subsection{Effect of Sensor Costs}
\begin{figure}[h]
\begin{center}
\includegraphics[width = 0.5\textwidth, height = 0.37\textwidth]{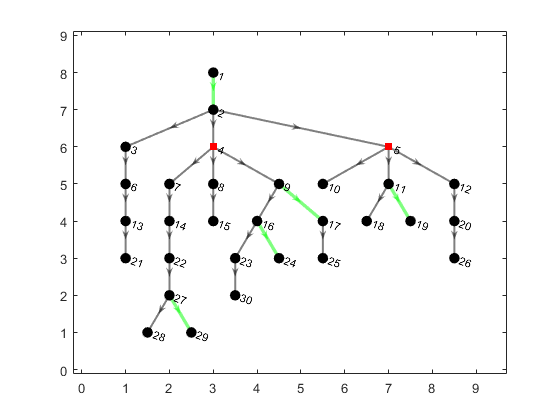}
\caption{TEST feeder: Case 1}\label{s1}
\end{center}
\end{figure}

\begin{figure}[h]
\begin{center}
\includegraphics[width = 0.5\textwidth, height = 0.37\textwidth]{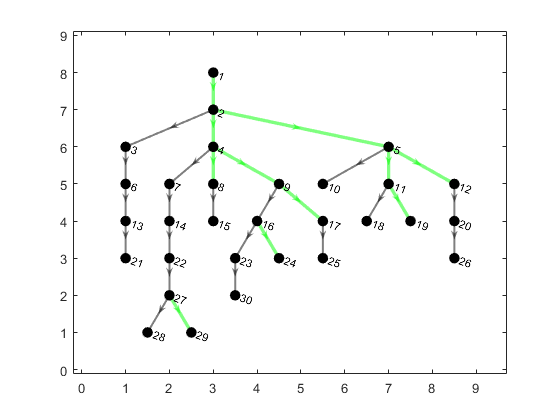}
\caption{TEST feeder: Case 2}\label{s2}
\end{center}
\end{figure}

In this simulation, we consider a single phase test feeder model (TEST) with 30 nodes and node $1$ as the root node. We find optimal sensor placement for this feeder under two different simulation cases. In the first case (case 1), we have $a_i = 2 \ \forall i \in V$, $b_{(i,j)} = 1 \ \forall (i,j) \in E$ and no zero-injection nodes in the network. The optimal sensor placement for case 1 is illustrated in Fig. \ref{s1}. In the second case (case 2) we have $a_i = 3 \ \forall i \in V$, $b_{(i,j)} = 1 \ \forall (i,j) \in E$ and again no zero-injection nodes. The optimal sensor placement for case 2 is illustrated in Fig. \ref{s2}. Table \ref{ts1} compares the two cases in terms of the number of node sensors $|V_{\Pcal}|$, number of line sensors $|E_{\Pcal}|$ and algorithm execution time. We can clearly observe that the placement algorithm does not place any node sensors in case 2 since the relative cost of a node sensor to a line sensor is higher in case 2 as compared to case 1. $|V_{eval}|$ being same for both the cases which makes the execution times similar.

\subsection{Presence of Zero-Injection Nodes}
Next we applied our sensor placement algorithm to modified versions of IEEE 37 and IEEE 123 feeders. The modifications performed are according to \cite{dobbe}. For both the feeders, the feeder voltage and power ratings are left unchanged. However, we assume that all lines are single phase, all loads in the network are constant power loads and ignore any transformers or voltage regulators that are present. We find an optimal sensor placement for each IEEE feeder under two different simulation cases: case 3 and case 4. In both case 3 and 4, we have $a_i = 2 \ \forall i \in V$, $b_{(i,j)} = 1 \ \forall (i,j) \in E$. However, in case 3 we have no zero-injection nodes while in case 4 we have $10$ zero-injection nodes for IEEE 37 feeder and $28$ zero-injection nodes for IEEE 123 feeder. Table \ref{ts1} compares the two cases for both IEEE 37 and IEEE 123 feeders in terms of the number of node sensors $|V_{\Pcal}|$, number of line sensors $|E_{\Pcal}|$ and algorithm execution time. From Table. \ref{ts1} it can be clearly observed that the presence of zero-injection nodes in the network has changed the optimal sensor placement. For both the IEEE feeders, the presence of zero-injection nodes under case 4 has increased the number of node sensors placed $|V_{\Pcal}|$. However, since $|V_{eval}|$ is almost same under both cases 3 and 4 for both IEEE feeders, the execution times are similar as can be seen from the Table. \ref{ts1}. 

\subsection{Scalability}
To illustrate the scalability of our dynamic programming algorithm we find optimal sensor placements under case 3 for a couple of networks of larger size. The first network is a $183$ node test feeder from our Department of Energy SHINES (Sustainable Holistic Integration of Energy Storage and Solar PV) project. The second network is a modified version of IEEE European low voltage test feeder with $906$ nodes. The modifications performed are same as those performed for IEEE 37 and 123 feeders. The results tabulated in Table \ref{ts1} illustrate that the execution time of our algorithm for these larger networks is still small, supporting the notion that our algorithm is scalable to larger networks. 

\begin{table}[h]
 \caption{Sensor placement results for various distribution feeders}
\label{ts1}
\begin{center}
 \begin{tabular}{| c | c | c |c | c | c |} 
\hline
Feeder & Case & $|V_{eval}|$ & $|V_{\Pcal}|$ & $|E_{\Pcal}|$ & Time ($10^{-4}$ s) \\
\hline
TEST & 1 & 7 & 2 & 5 & 4.7 \\
\hline
TEST & 2 & 7 & 0 & 11 & 4.4 \\
\hline
IEEE 37 & 3 & 12 & 1 & 12 & 7.8 \\
\hline
IEEE 37 &  4 & 13 & 7 & 5 & 7.7 \\
\hline
IEEE 123 & 3 & 34 & 4 & 31 & 16 \\
\hline
IEEE 123 & 4 & 35 & 18 & 12 & 15 \\
\hline
SHINES & 3 & 62 & 1 & 61 & 35 \\
\hline
European & 3 & 97 & 7 & 86 & 70 \\
\hline
\end{tabular}
\end{center}
\end{table}

%% file: section7.tex
\section{Conclusions}\label{section7}

In this paper, we have presented a dynamic programming based minimum cost sensor placement solution for outage detection. We have introduced systematic definitions of concepts such as topology detectability and outage identifiability that are important in the context of outage detection. Then, we have formulated the sensor placement as a cost optimization problem with constraints that are necessary and sufficient for outage identifiability. Given the assumed tree structure of the networks, we proposed a dynamic programming algorithm that solves the sensor placement problem in polynomial time. Finally, we have presented numerical results that illustrate the properties of our sensor placement algorithm for multiple feeder models. Currently, we are developing distributed outage detection algorithms that utilize the proposed sensor placement solution to detect line outages in a distributed manner. Also, we are working on extending our placement solution, and developing outage detection algorithms for  distribution networks with {distributed generators (DGs) and grid-connected} microgrids since these networks raise unique challenges \cite{telu} that were not considered in the current formulation. 